\documentclass[draft]{agujournal2019}
\usepackage{url} 
\usepackage{lineno}
\usepackage[inline]{trackchanges} 
\usepackage{soul}

\catcode`\@=11
\newcommand{\gsim}{${\mathrel{\mathpalette\@versim>}}$}
\newcommand{\lsim}{${\mathrel{\mathpalette\@versim<}}$}
\newcommand{\@versim}[2]{\lower 2.9truept \vbox{\baselineskip 0pt \lineskip
    0.5truept \ialign{$\m@th#1\hfil##\hfil$\crcr#2\crcr\sim\crcr}}}
\catcode`\@=12

\newcommand{\ddeg}{$^{o}$}
\newcommand{\kms}{km s$^{-1}$}

\def\be{\begin{equation}}
\def\bea{\begin{eqnarray}}
\def\ee{\end{equation}}
\def\eea{\end{eqnarray}}
\newcommand{\arcmin}{$^{'}$}

\draftfalse

\journalname{Radio Science}

\begin{document}

\title{A Cryogenic Wideband (2.5 - 14 GHz)
 Receiver system for the Arecibo 12m Telescope}

%
%




\authors{D. Anish Roshi\affil{1,2}, Phil Perillat\affil{1}, Felix Fernandez\affil{1}, Hamdi Mani\affil{3}, Benetge Perera\affil{1,2}, Periasamy~K.~Manoharan\affil{1,2}, Luis Quintero\affil{1}, Arun Venkataraman\affil{1}}


\affiliation{1}{Arecibo Observatory, Arecibo, Puerto Rico 00612}
\affiliation{2}{University of Central Florida, 4000 Central Florida Blvd, Orlando, FL 32816}
\affiliation{3}{CryoElec LLC, Chandler, Arizona}




\correspondingauthor{D. Anish Roshi}{anish.roshi@gmail.com}




\begin{keypoints}
\item Successful commissioning of a wideband, cryogenic receiver for the 12m telescope at the Arecibo Observatory.
\item Achieved system temperature of 40 K and gains of 0.025 K/Jy and 0.018 K/Jy  at frequencies 3.1 and 8.6 GHz respectively.
\end{keypoints}

%
%

%
%


\begin{abstract}
In this paper we present details of the construction of a wideband, cryogenic receiver and its successful commissioning on the Arecibo 12m telescope.  The cryogenic receiver works in the 2.5-14 GHz frequency range. The telescope is operated by the Arecibo Observatory, and is located within the premises of the Observatory. We upgraded the current narrow band, room temperature receivers of the telescope with the new wideband receiver. The current receiver is built around a Quadruple-Ridged Flared Horn (QRHF) developed by \citeA{Akgiray13}. To mitigate strong radio frequency interference (RFI) below 2.7 GHz, we installed a highpass filter before the first stage low noise amplifier (LNA). The QRHF, highpass filter, noise coupler and LNA are located inside a cryostat and are cooled to 15 K. The measured receiver temperature is 25 K (median value) over 2.5 GHz to 14 GHz. The system temperature measured at zenith is about 40 K near 3.1 and 8.6 GHz and the zenith antenna gains are 0.025 and 0.018 K/Jy at the two frequencies respectively. In the next stage of the development, we plan to upgrade the highpass filter in order to achieve  better RFI rejection near 2.5 GHz, improve the aperture efficiency at 8.6 GHz and upgrade the IF system to increase the upper frequency of operation from 12 GHz to 14 GHz.
\end{abstract}

\section{Introduction}

The Arecibo Observatory (AO), in addition to hosting the 305m legacy telescope, had a 12 m diameter dish antenna. The telescope is located within the AO site (longitude: 66\ddeg\ 45\arcmin\ West, latitude: 18\ddeg\ 20\arcmin, altitude: 496m). The 12m telescope was commissioned in 2011 with room temperature receivers operating at S (2.21 -- 2.34 GHz) and X (8.1 -- 9.2 GHz) bands, but it was not widely used for astronomy observations. In 2022, after the collapse of the 305 m telescope, the receivers of the 12m telescope were integrated with the legacy telescope backends. The Mock spectrometer was used for testing and later for observations. We also used the telescope along with the Very Long Baseline Interferometer (VLBI) backend to make successful fringe test observations with the European VLBI network. The system temperatures at S and X bands, measured after integration, were respectively 104 K (system equivalent flux density, SEFD, 2895 Jy) and 125 K (SEFD 3459 Jy). Despite its high system temperatures, the 12m telescope was extensively used for Solar radio continuum imaging at X band\cite{mano23}, high-cadence Pulsar monitoring - more than 20 bright pulsars were monitored \footnote{\url{https://www.naic.edu/~phil/hardware/12meter/pulsars/pulsars.html}} \cite{per23} and FRB studies \cite{ben22}.  We also studied feasibility of measuring long term variability of quasar flux density with the 12m telescope. 

The higher system temperature and relatively narrow bandwidth of operation are not ideal for wider scientific use of the telescope. After discussions with the scientific user community, we decided to upgrade the current frontend with a wideband (2.5 to 14 GHz), cryogenic receiver. Since only one feed could be placed in the feed housing, we decided to commission a wideband receiver that would allow the use of the 12m telescope for a large range of science projects. A brief description of the 12m telescope and feed housing is given in Section~\ref{sec:12m}. The upgraded receiver system and backend are described in Section~\ref{sec:wbrec} and the measured performance of the new receiver on the telescope is given in Section~\ref{sec:sysper}. The results from commissioning observations are briefly discussed in Section~\ref{sec:com}, which demonstrate the range of science projects that could be done with the new system. Finally, the plans for future development work are presented in Section~\ref{sec:future}, which is followed by a summary (Section~\ref{sec:sum}) of the paper. 
 
\section{The 12m telescope and feed housing}
\label{sec:12m}

\begin{figure*}[ht!]
    \centering
\begin{minipage}{.7\textwidth}
  \centering
  \includegraphics[trim={30cm 0 0 0},clip,width=\textwidth, height=8cm]{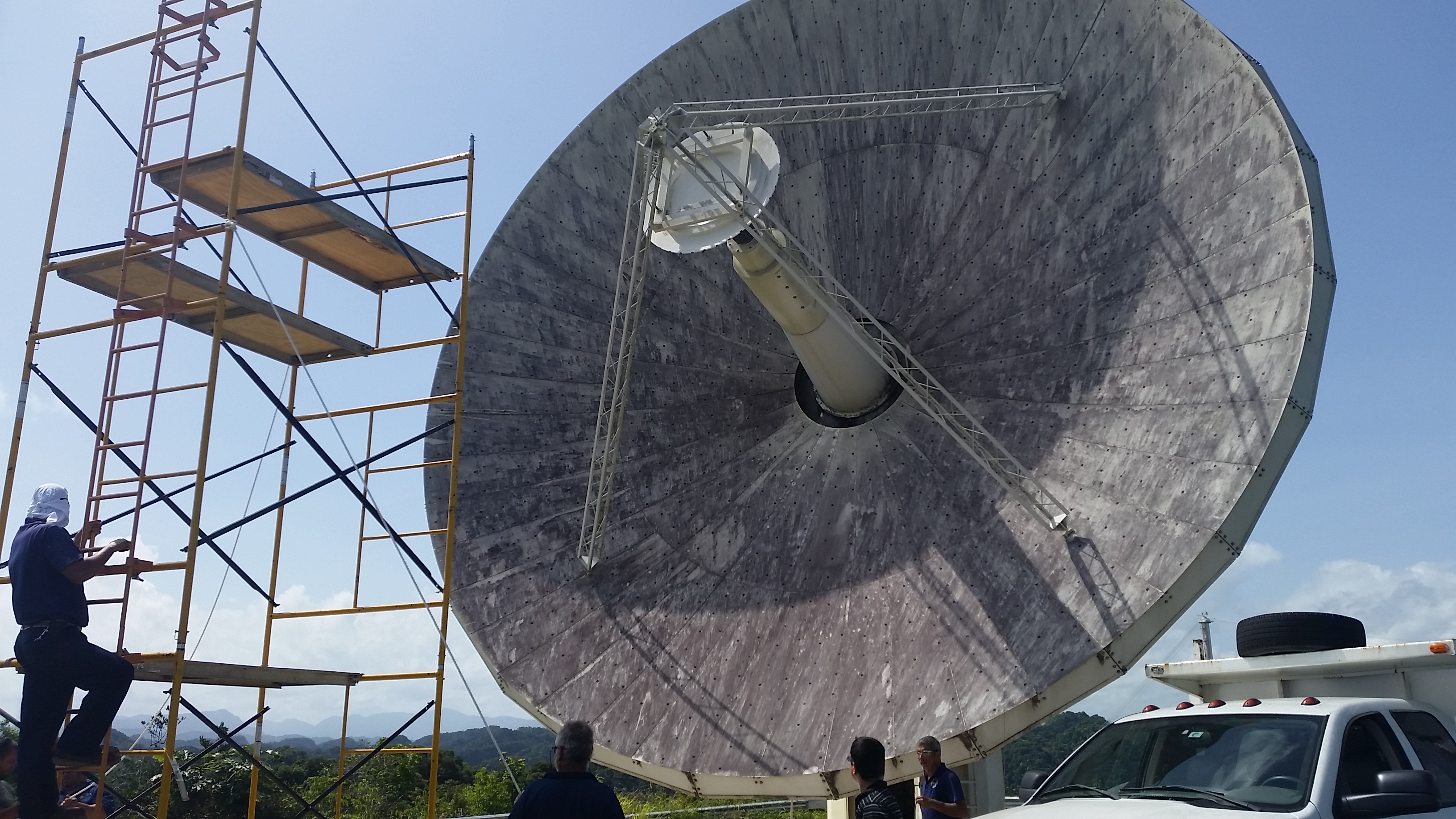}
\end{minipage}%
\begin{minipage}{.3\textwidth}
  \centering
  \includegraphics[width=\textwidth, height=8cm]{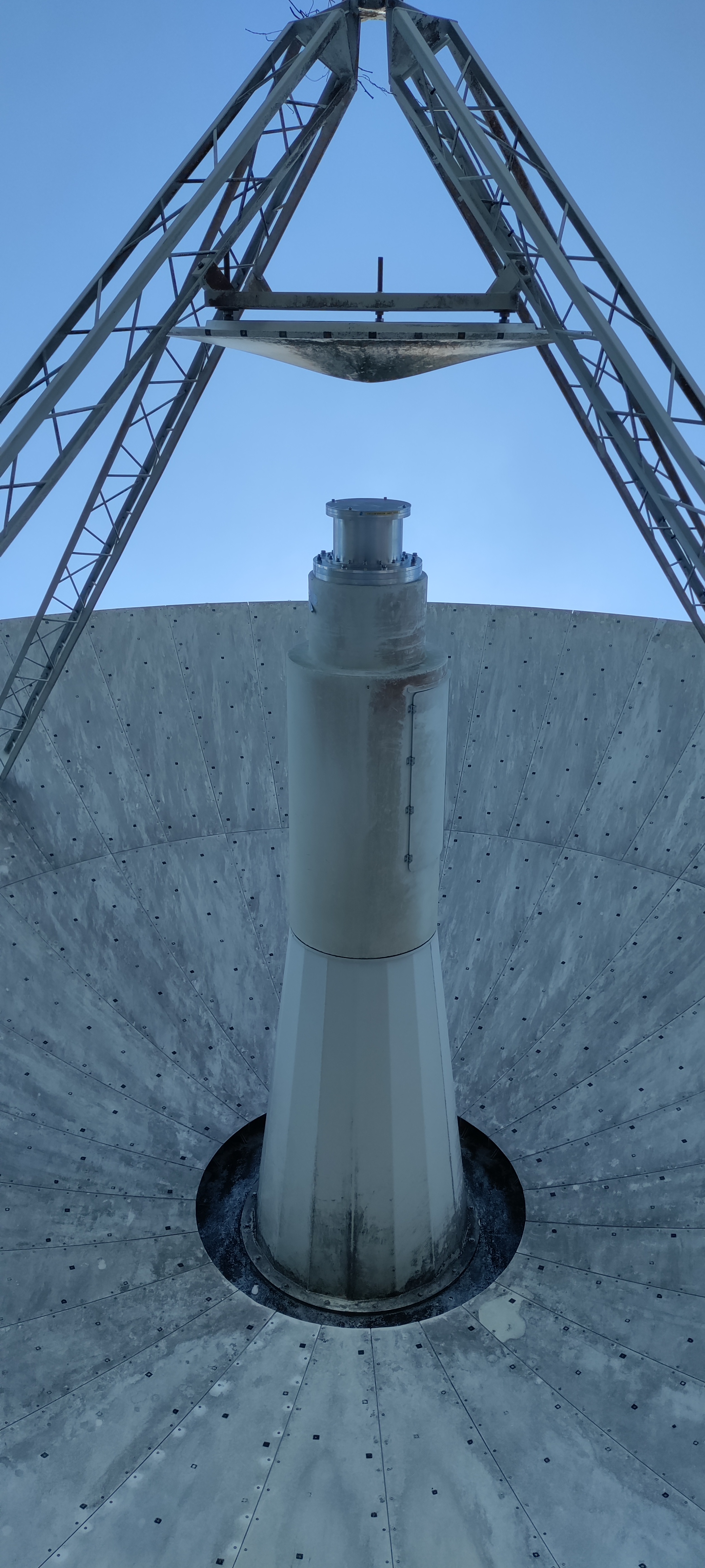}
\end{minipage}
\caption{{\bf Left:} The 12m radio telescope. The telescope is located within the Arecibo Observatory site. {\bf Right:} The new wideband (2.5-14 GHz), cyrogenic receiver is mounted on the feed cone extension. The image shows the feed installation with two shims, but after focus adjustment, 4 shims each of 1 inch thickness were used to position the feed at the focus (see text).}
\label{fig:fig1}
\end{figure*}

The 12m telescope is a fully steerable reflector antenna (see Fig.~\ref{fig:fig1}). The alt-azimuth mount of the telescope allows  an elevation range 5.8\ddeg\ to 88\ddeg\ and azimuth range $-$180\ddeg\ to 360\ddeg. The maximum slew rates and accelerations of the telescope are 5.0\ddeg\ s$^{-1}$ and 1.33\ddeg\ s$^{-2}$ in azimuth and  1.2\ddeg\ s$^{-1}$ and 2.6\ddeg\ s$^{-2}$ in elevation. 

The optical configuration of the 12m telescope is a Cassegrain reflector system. The primary reflector is made up of stretched form panels. They are supported by a backup structure consisting of a rigid hub, support ribs and interconnecting bracing. During the site acceptance test, the measured surface RMS was 381 $\mu$m and it remained more or less the same over 10\ddeg\ to 88\ddeg\ elevation range. The prime focal length over diameter (F/D) ratio of the telescope is 0.375. The sub-reflector is fabricated using carbon fiber and is attached to a quad support structure.

The feed is located in front of the sub-reflector, mounted in the feed cone extension (see Fig.~\ref{fig:fig1}). The feed cone extension is fixed on the face of the rigid hub that forms a part of the backup structure of the main reflector. The feed cone extension is painted white to reduce the thermal gradient. The new cryogenic system was designed such that the cryostat and a subset of room temperature electronics could be mounted inside the feed cone extension, the latter are accessible through an access panel.  

\section{The New Wideband Feed, Receiver System, Backends and Data Processing}
\label{sec:wbrec}

\begin{figure*}[ht!]
    \centering
\begin{minipage}{.5\textwidth}
  \centering
  \includegraphics[height=5.2cm, width=\textwidth]{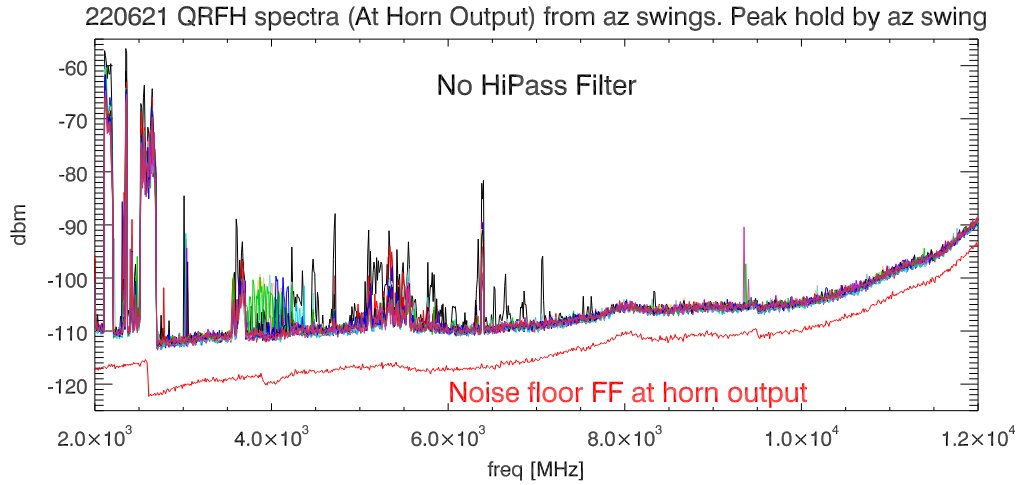}
\end{minipage}%
\begin{minipage}{.5\textwidth}
  \centering
  \includegraphics[height=5cm, width=\textwidth]{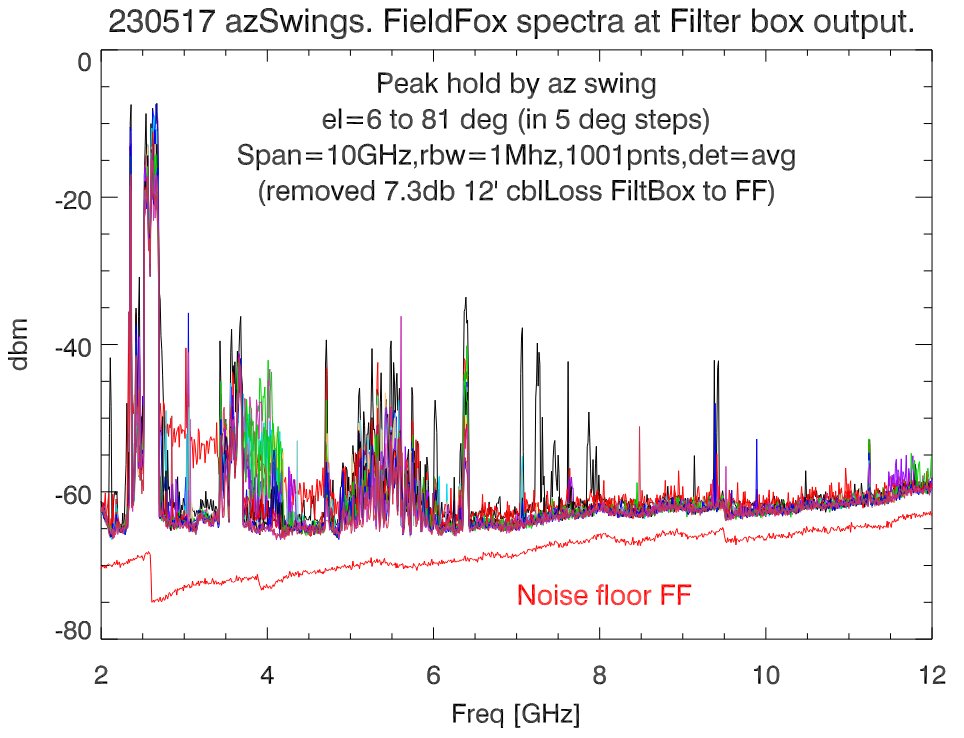}
\end{minipage}
\caption{{\bf Left:} Measured radio spectra at the output of QRFH. The QRFH with a warm amplifier was mounted on the feed cone extension without the cryostat for these measurements (see text). {\bf Right:} Spectra measured with the cryogenic receiver mounted on the feed cone extension. The spectra are obtained at the output of the filter box after selecting the through path in a similar fashion as shown in the left panel (see Section~\ref{sec:future}). The power values can be translated to the output of the QRFH by subtracting about 48 dB. These measurements were made with the Fieldfox spectrum analyzer and the noise floor of the analyzer is shown in red curve. }
\label{fig:fig2}
\end{figure*}

\begin{figure*}[ht!]
    \centering
\begin{minipage}{.5\textwidth}
  \centering
  \includegraphics[trim={10cm 20cm 30cm 10cm},clip, width=\textwidth, angle=-90]{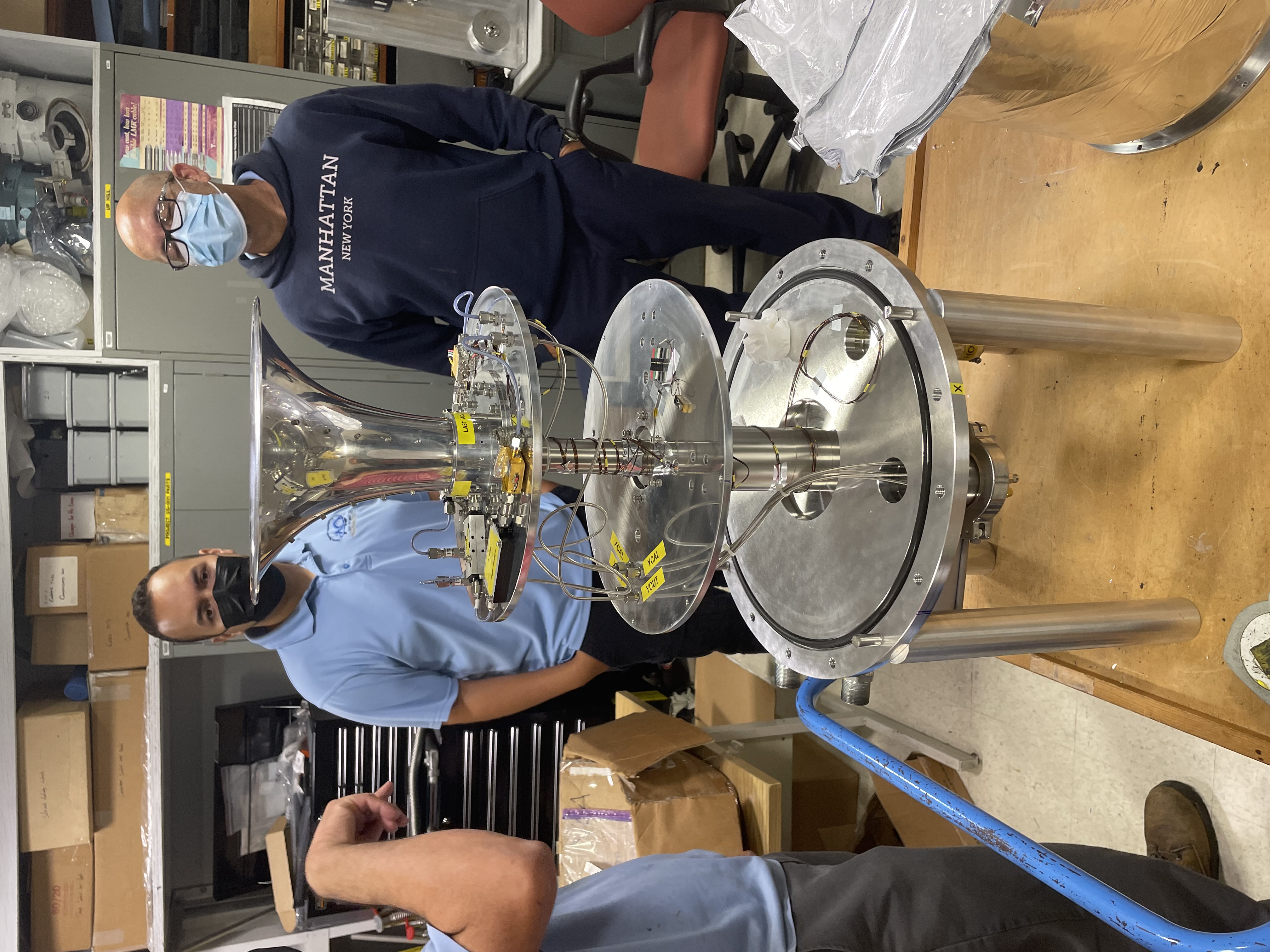}
\end{minipage}%
\begin{minipage}{.5\textwidth}
  \centering
  \includegraphics[width=\textwidth]{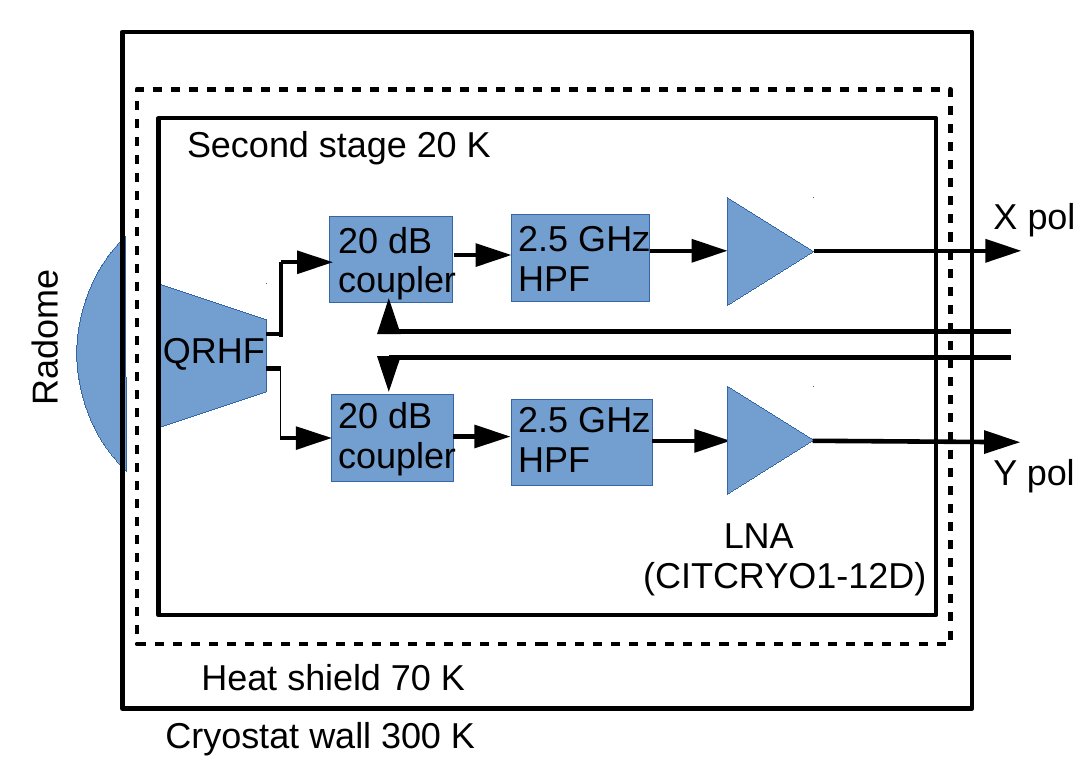}
\end{minipage}
\caption{{\bf Left:} The QRHF and the electronic components inside the cryostat. {\bf Right:} The block diagram showing the cryostat stages, components at 20 K stage and the Radome.}
\label{fig:fig3}
\end{figure*}

Only one feed can be mounted on the feed cone extension. In order to maximize the scientific usability of the telescope, we decided to upgrade the 12m telescope receiver with a wideband, cryogenic receiver. In 2013, \citeA{Akgiray13}  developed a circular Quadruple-Ridged Flared Horn (QRHF) that can achieve an almost constant beamwidth over 6:1 bandwidth. A slightly modified version of QRFH, which could cover frequency range $\sim$ 2.3 to 14 GHz was selected for our application. The feed and the cryostat was designed and built by CryoElec, LLC, Arizona, in consultation with AO engineers and scientists. 

The 12m telescope is located in the Arecibo Observatory radio coordination zone. In spite of its location, it is important to characterize radio frequency interference (RFI) when designing a wideband receiver. Fig.~\ref{fig:fig2} (left) shows the radio spectra measured over the 
frequency range $\sim$ 2 GHz to 14 GHz. These measurements were made with the QRFH feed along with a warm amplifier mounted on the feed extension cone without the cryostat. The 12m antenna was moved in azimuth (360\ddeg\ span) and elevation (6\ddeg\ to 81\ddeg\ in 5\ddeg\ step) during the measurements. Each spectrum corresponds to a full Azimuth swing at constant elevation. All measurements were made with a Keysight Fieldfox spectrum analyzer (N9916A\footnote{\url{https://www.keysight.com/us/en/assets/7018-03314/data-sheets/5990-9783.pdf}}) with resolution bandwidth set to 1 MHz. The noise floor of the analyzer is shown in red curve in Fig.~\ref{fig:fig2}.  The system gain was estimated and the measured spectra were divided by the gain so that the power levels in the figure represent values at the output of the QRFH. The power values shown are peak values; the average values are roughly 10 dB below the peak values. As seen from these plots, there are several strong RFI (\gsim\ -80 dBm) present at frequencies below 2.7 GHz.   

\begin{figure*}[ht!]
\centering
\includegraphics[width=\textwidth]{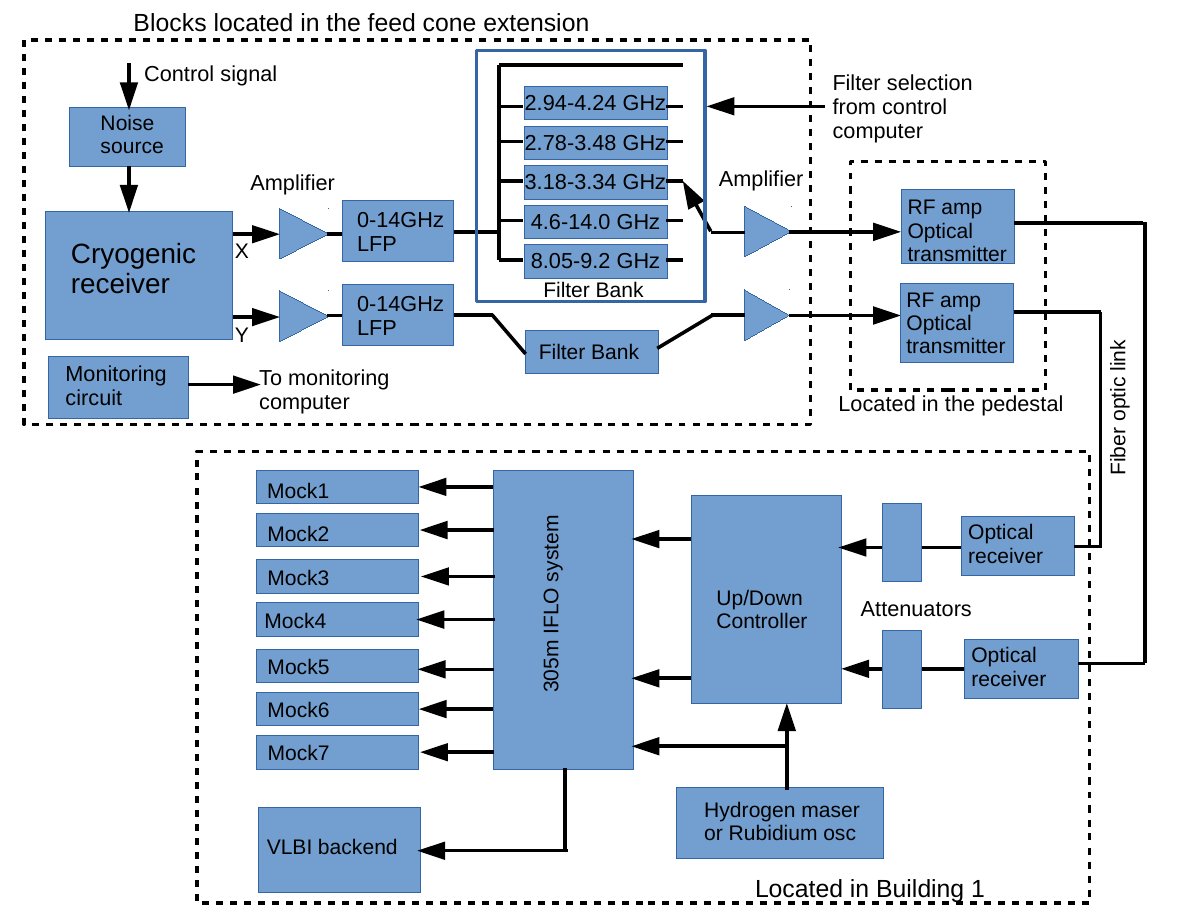}
\caption{A block diagram of the 12m telescope cryogenic receiver and backends}
\label{fig:fig4}
\end{figure*}

We selected the cryogenic low noise amplifier (LNA) Citcryo1-12D\footnote{\url{https://www.cosmicmicrowavetechnology.com/cit112}} for our application. This amplifier has a gain of 35 dB and noise temperature \lsim\ 7 K over 2.5 to 12 GHz ($\le$ 10 K up to 14 GHz) when cooled to a physical temperature of 12 K. The input 1 dB compression point is -35 dBm. Our estimate of the total integrated RFI power with a 2.5 GHz highpass filter (rejection at 2.37 GHz is -20 dB) is about -65 dBm, which gives an operating dynamic range $\sim$ 20 dB. We define the operating dynamic range based on the consideration that the non-linearity of the amplifier is not severe for power levels up to -45 dBm (i.e 10 dB below the 1 dB compression point). Based on these estimates, we decided to have a 2.5 GHz highpass filter before the LNA.

Fig.~\ref{fig:fig3} (left) shows the assembled QRFH and some of the components placed inside the cryostat. The measurements made on the QRFH showed excellent return loss ($\sim$ -10 db) and cross polarization performance ($\le$ -30 dB) over the 2.5 to 14 GHz frequency range. The native polarization of the feed is linear, referred to as X and Y polarizations in this paper. The cryostat was built around a two stage Gifford-MacMohan (GM) cryocooler. We are using the commercial Trillium 350CS cold head\footnote{\url{https://trilliumus.com/products/cryogenics/cryo-plex-cryo-pumps/cryogenic-cold-heads/}} and M600 Helium compressor\footnote{\url{https://trilliumus.com/products/cryogenics/m-series-compressors/trillium-m600-cryo-compressor/}} to implement the GM cryocooler. A block diagram of the components inside the cryostat is shown in Fig.~\ref{fig:fig3} (right). The cryostat encloses the QRHF, noise couplers, 2.5 GHz high-pass filters and the low-noise amplifiers. Due to the relatively high humidity at Arecibo, we built a radome with Mylar material in front of the cryostat radio frequency window. Warm, compressed air is circulated within the radome to mitigate condensation. A commercial air dryer (Ingersoll Rand Refrigerated Air Dryer 11CFM Model D18IN) and compressor (Ridgid 6 Gallon Portable Electric Pancake Air Compressor Model OF60150HB) were used for this purpose. The cryostat window is also made of Mylar material. In addition to the blocks shown in Fig.~\ref{fig:fig3} (right), we have installed diode temperature sensors inside the cryostat to monitor the temperature of the two stages, and sensors to  monitor the gate, drain, supply voltages and the drain current of the LNAs. A vacuum gauge was used to report the pressure inside the cryostat. The helium supply and return pressures in the compressor were also monitored. 

A block diagram of the new receiver system and its interface to the backends is shown in Fig.~\ref{fig:fig4}. The signals from the X and Y polarization outputs of the cooled receiver are further amplified and filtered using a 0 to 14 GHz low pass filter. They are then passed through a filter bank consisting of a through path providing the full 2.5 to 14 GHz bandwidth and 5 separate bandpass filters. The bandpass filters for the observations are selected from a remote control computer. All these components are located inside the feed cone extension. The outputs of the filter bank are amplified and then connected to an analog fiber optic transmitter, which is located at the antenna pedestal.

 The radio frequency (RF) modulation on the optical signals are received using optical receivers and converted back to RF signals at Building 1 at AO. These signals are passed through fixed attenuators and connected to the `Up/down convertor' (UDC) module. The local oscillators (LO) in the module can currently select a 1 GHz bandwidth anywhere in the 2.5 to 12 GHz frequency range  (see Section~\ref{sec:future} for future upgrade needed to increase the coverage up to 14 GHz). This 1 GHz bandwidth signal from a polarization is centered at 1.5 GHz intermediate frequency (IF) and is connected to the `IFLO system' of the legacy 305 m telescope. The frequency of the signals are translated and bandwidths limited in the IFLO system such that the seven Mock spectrometers can flexibly be tuned to subbands within the 1 GHz bandwidth. A separate interconnection from the IFLO system is available to the use the Roach Digital Backend (RDBE) and the Mark 6 recorders for VLBI observations.

Most observations during commissioning were made with the Mock spectrometers. The maximum bandwidth of operation of each Mock spectrometer is about 172 MHz, and the maximum spectral channels provided by each spectrometer is 8192. Thus the seven Mock spectrometer can cover a maximum of $\sim$ 1.0 GHz (assuming 10\% overlap on either side of the 172 MHz band) with a spectral resolution of 21 KHz in full polarization mode. For spectroscopic observations that require higher resolution and lower bandwidth per spectrometer, digital filters are available in the Mock spectrometer to reduce the bandwidth. The number of channels can also be varied between 64 and 8192. The measured spectra can be calibrated in antenna temperature by acquiring data with the calibration noise source at the front-end turned on and off in separate scans. The same spectrometer can be used for pulsar observations; the minimum integration time available is 32 $\mu$s. The Mock spectrometer is also capable of switching the noise calibration diode in the front-end at 25 Hz rate, which is useful for calibrating pulsar observations. Currently all the non-pulsar data processing is done using IDL routines developed by Phil Perillat\footnote{\url{http://www.naic.edu/~phil/software/software.html#idldoc}} and pulsar data processing is done using the pipeline \texttt{AO12PSR}\footnote{\url{https://github.com/benperera/ao12psr}} developed by Benetge Perera.

\section{System Performance}
\label{sec:sysper}

\begin{figure*}[ht!]
\centering
\includegraphics[height=0.3\textheight, width=\textwidth, angle=0]{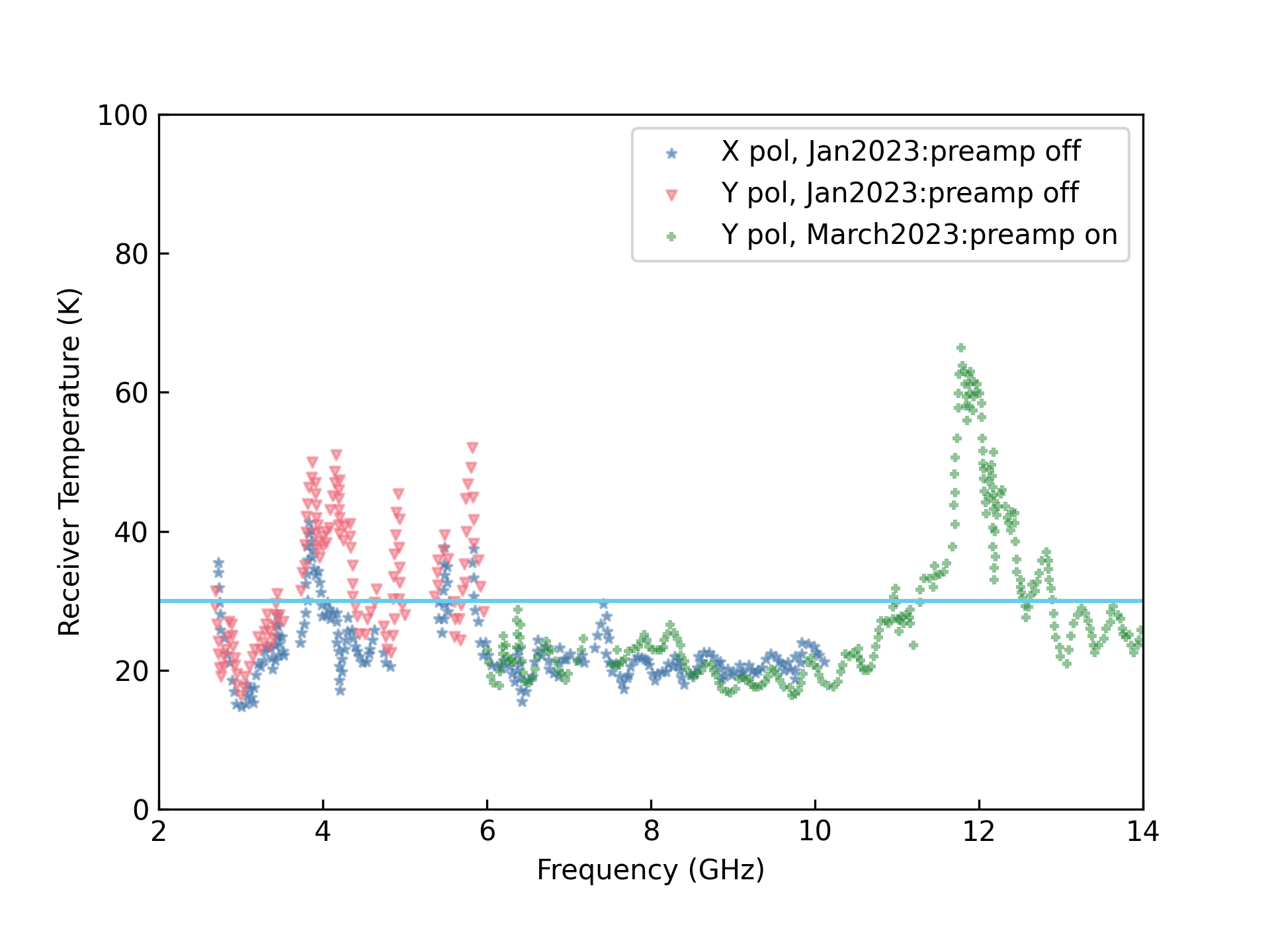}
\caption{Receiver temperature of the cryogenic system as a function of frequency. The measurements were made at the outdoor antenna test building at AO. The receiver temperature for X polarization is shown for frequencies from 2.5 to 10 GHz. Two sets of measurements were made with signals from the Y polarization each of them covers frequencies 2.5 to 6 GHz and 6 to 14 GHz (see text). The horizontal line shows the specified 30 K temperature. The excess temperatures above 30 K at a few frequencies below 6 GHz and near 12 GHz are respectively due to terrestrial and satellite RFI affecting the measurements. Also measurements severally affected by RFI are masked out in the plot.}
\label{fig:fig5}
\end{figure*}

\begin{figure*}[ht!]
\centering
\begin{minipage}{.5\textwidth}
  \centering
\includegraphics[width=\textwidth]{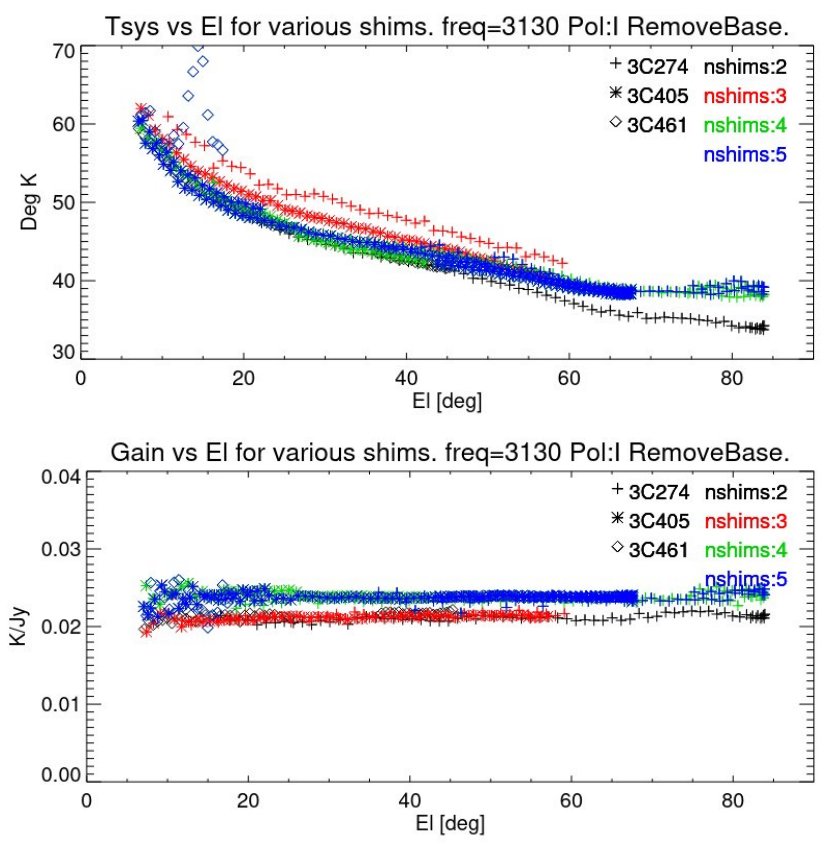}
\end{minipage}%
\begin{minipage}{.5\textwidth}
  \centering
\includegraphics[width=\textwidth]{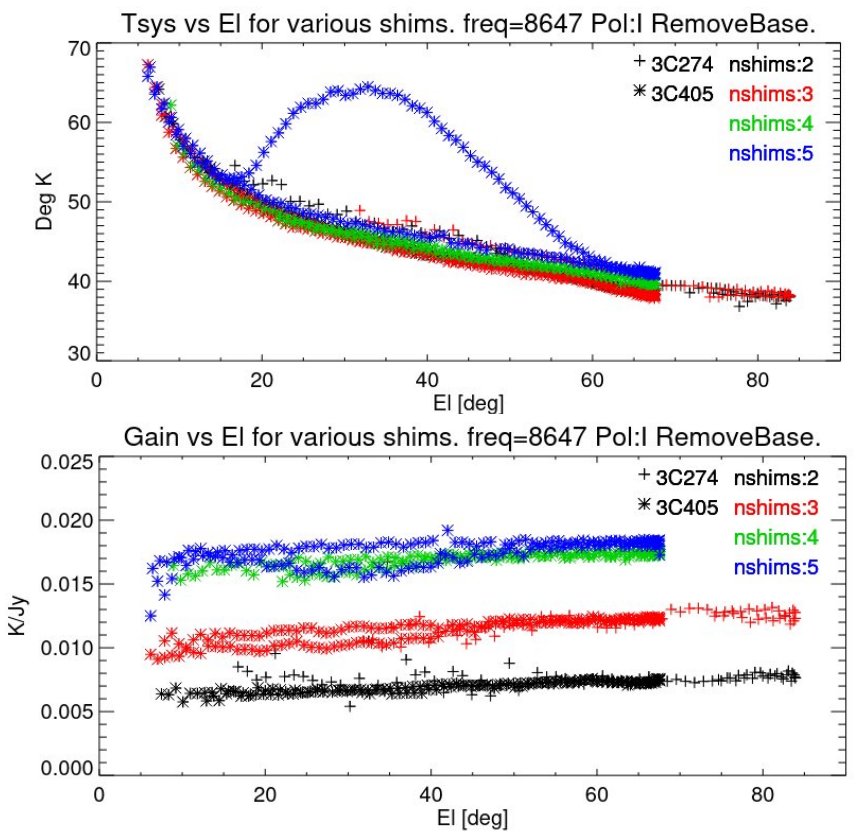}
\end{minipage}
\caption{{\bf Left:} The measured system temperature and telescope gain in K/Jy as a function of elevation near 3.1 GHz. {\bf Right:} The same quantities for 8.6 GHz. The current feed position corresponds to nshim=4. The excess system temperature in the elevation range $\sim$20-60\ddeg\ for nshim=5 is due to bad weather during the measurements.}
\label{fig:fig6}
\end{figure*}

Receiver temperature of the cryogenic system was measured at the outdoor antenna test facility using the Y-factor method. An absorber placed in front of the feed horn formed the `hot load' and sky emission formed the `cold load'.  We used 295 K for the hot load. The cold load temperature needs to include contribution from the atmosphere; we used a model developed by P. Goldsmith\footnote{\url{http://www.naic.edu/~phil/temp/pfg_obs_opt_depth.ps}}. The zenith sky temperature range from 5 K to 14 K for frequencies 1 to 10 GHz respectively; these values were used for the `cold load' temperature. The power spectrum measurements to estimate the Y-factor were made with fieldfox spectrum analyzer (N9916A). The measured receiver temperature for the two polarizations are shown in Fig.~\ref{fig:fig5}. The measurements were done with preamplifier of the spectrum analyzer in off state for frequencies up to 6 GHz due to RFI. In order not to degrade the sensitivity, the preamplfier had to be turned on for frequencies above 6 GHz. In Fig.~\ref{fig:fig5} we show the two set of measurements for Y polarization. For X polarization, the measurements were made with preamplfier in off state up to 10 GHz. The feed was pointing to zenith when these measurements were made. Measurements at several frequencies below 6 GHz were affected by RFI, which are mostly masked in the plot (but some of the temperature value above 30 K are affected by RFI). The measurements near 12 GHz are affected by RFI from satellite transmission, which has resulted in system temperature estimation above 30 K. Discarding these RFI affected measurements, the receiver temperature is 25 K (median value), which is better that our specified value for the receiver construction.

\begin{table}
 \caption{Characteristics of the Cyrogenic Receiver system}
 \centering
 \begin{tabular}{c c c c c c c c}
 \hline
  Frequency & Trec$^a$ & Tsky$^b$ & Tsys$^c$ & Tscat$^d$ & S$_{cygA}^e$ & Gain$^f$  & $\eta^g$ \\
  (GHz)     &  (K) & (K) &  (K) &  (K)   & (Jy)       & (K/Jy) &  (\%)      \\
 \hline
   3.1      & 22  &  5.5  & 40   &  12.5 & 643        & 0.025  & 61 \\
   8.6      & 21  &  7.5  & 40   &  11.5 & 185        & 0.018  & 44 \\
 \hline
 \multicolumn{8}{l}{$^a$Measured receiver temperature.}\\
 \multicolumn{8}{l}{$^b$Estimated sky temperature (see text)}\\
 \multicolumn{8}{l}{$^c$Measured system temperature.}\\
 \multicolumn{8}{l}{$^d$Estimated scatter temperature (see text).}\\
 \multicolumn{8}{l}{$^e$Flux density of Cygnus A used for estimation.}\\
 \multicolumn{8}{l}{$^f$Measured telescope gain}\\
 \multicolumn{8}{l}{$^g$Estimated aperture efficiency}
 \end{tabular}
 \label{tab:char}
 \end{table}
 
The cryogenic system was installed on the telescope in May 2023. The system power levels were initially adjusted by inserting fixed attenuators along the receiver chain in order to maintain linearity up to the digitizers in the Mock spectrometer. These adjustments were made for the two selected filters 2.78-3.48 GHz and 8.05-9.2 GHz. The temperature of the noise source was then calibrated using the Y-factor method by measuring the power with an absorber placed in front of the feed while mounted on the telescope and on cold sky. We then obtained cross scan measurements with the two filters on three calibrators - Cygnus A (3C405), Cas A (3C461) and M87 (3C274). These scans were made along azimuth at constant elevation and along elevation spanning about 5 full width at half maximum (FWHM) beamwidths on either side of the source. At the beginning of each scan, the calibration noise was injected, which is used to estimate the off-source system temperature. The deflection due to the source along with the flux densities of the sources were used to estimate the gain of the telescope in units of K/Jy. These quantities are plotted in Fig.~\ref{fig:fig6} as a function of elevation. The measurements were repeated for different distances of the feed from the secondary reflector (which are marked as nshim 2 to 5 in Fig.~\ref{fig:fig6}) to determine the optimum focal position. We varied the distance from the apex of the secondary reflector to the face (or mouth) of the horn between 24.6 and 19.6 inch in steps of 1 inch. The distance which provided the maximum gain over system temperature is 20.6 inch (corresponding to 4 shims). The system temperatures estimated at zenith are about 40 K at 3.1 and 8.6 GHz and zenith gains are 0.025 K/Jy and 0.018 K/Jy near 3.1 GHz and 8.6 GHz respectively. Table~\ref{tab:char} summarizes the measured and estimated parameters. The scatter temperature listed in Table~\ref{tab:char} is the spillover temperature plus contribution from any additional scattered radiation. The flux density of Cygnus A used for estimation is also listed, which is obtained from the model of \citeA{pb17}.

We also used the cross scan measurements to obtain the FWHM beamwidth and pointing errors. The FWMH beamwidths for the average power of the two polarizations are about 27\arcmin and 11\arcmin respectively for the 3.1 and 8.6 GHz frequencies. The ratio of the beamwidths measured along azimuth and elevation is close to unity. Our measurements show that the ratio of the beamwidths for the X and Y polarizations are 1.1 and 0.9 respectively at 8.6 GHz, indicating the beam cross section for each polarization is elliptical. Such ellipticity for the individual polarization beams is not present at 3.1 GHz. The measured pointing errors relative to the model developed for the previous room temperature receiver are \lsim\ 1.5\arcmin\ for elevation range 10 to 88 \ddeg\footnote{Plot for these measurements as well as results for different frequencies are available at \url{http://www.naic.edu/~phil/hardware/12meter/wbrcvr/commission/shims/wbshims.html}. } 

\section{Results from Commissioning Observations}
\label{sec:com}

\begin{figure*}[ht!]
\centering
\includegraphics[height=0.7\textheight, width=0.9\textwidth]{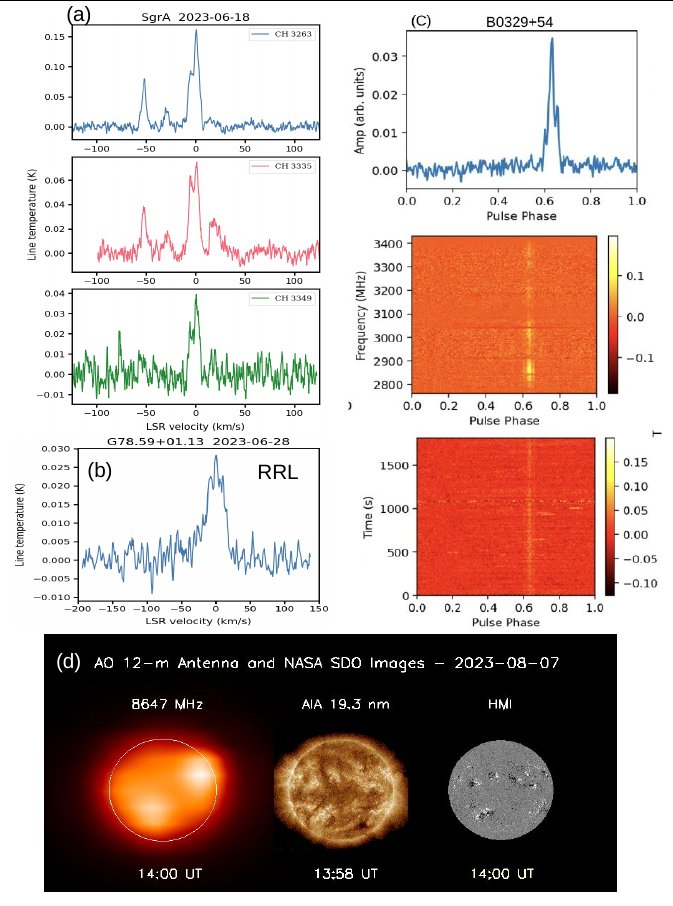}
\caption{{\bf (a)} The 3.263, 3.335, 3.349 GHz transitions of CH (Methylidyne) radical toward the Galactic center (see text). {\bf (b)} Hydrogen radio recombination line (RRL) observed toward Cygnus X region (G78.59+1.13) near 8.6 GHz. {\bf (c)} Pulsar B0329+54 observed near 3.1 GHz. The average pulse profile is shown on the top, frequency vs pulse phase in the middle and pulse amplitudes vs pulse phase in the bottom. {\bf (d)} Image of the Sun (left) at 8647 MHz made on 07 August 2023 with the cryogenic, wideband receiver of the 12m telescope along with near-simultaneous EUV image of the Sun at 19.3 nm (middle) and the photospheric magnetogram (right).}
\label{fig:fig7}
\end{figure*}

A number of commissioning observations with the cryogenic receiver were done mid June, 2023. Fig.~\ref{fig:fig7} shows some of the results from these observations. The hyperfine structure of the ground state of CH (Methylidyne) radical results in three transitions of frequencies 3.263, 3.335 and 3.349 GHz. The 12m telescope was used to observe these transitions toward the Galactic center. We selected the filter 3.18 to 3.34 GHz filter for these observations. Three Mock spectrometers were used to observe the three transitions in on-source only observing mode for commissioning. We used bandwidth per spectrometer of 24.6 MHz and 8192 spectral channels to get about 0.3 \kms spectral resolution. The bandpasses were approximated with a combination of harmonic and polynomial functions. The residual spectral baseline after bandpass correction were removed by a 7th order polynomial. The final spectra are shown in Fig.~\ref{fig:fig7}a. The amplitudes of the spectral features are likely to be affected by the bandpass and spectral baseline removal process adopted here. Subsequently we have developed a `pseudo on-off' observing mode for bandpass calibration of extended line emission like the CH line emission in the galactic plane. In this mode, an off-source position several degrees away is observed every 5 minutes. As discussed below, this method of bandpass calibration has resulted in residual spectral baseline that could be removed by polynomial of order 3 or less. We plan to survey CH line emission in the galactic plane using this mode of observation. CH line is known to be a good tracer for low-density (10${^2}$ - 10$^{3}$ cm$^{-3}$) molecular gas which are not traceable using CO emission - the so called CO dark gas \cite{Magnani05}. So far a large scale survey of CH line emission from the galactic plane does not exist and the 12m telescope will be ideal for such a survey.

The radio recombination line detection near 8.6 GHz toward the position G78.59+1.13 in Cygnus X region is shown in Fig.~\ref{fig:fig7}b. We selected the 8.05-9.2 GHz filter for these observations. Four recombination line transitions (H89$\alpha$, H90$\alpha$, H91$\alpha$ and H92$\alpha$) were simultaneously observed with 4 Mock spectrometers in a `pseudo on-off' observing mode.  Diffuse ionized gas is present over several degrees in Cygnus X region and we are interested in detecting lines from such extended ionized gas. Therefore an off-source position several degrees away (the off-source in this case G80.7+8.1) was observed every 5 minutes. The bandwidth and the number of spectral channels selected for Mock spectrometer were 57.3 MHz and 2048 respectively, which will provide $\sim$ 1\kms\ spectral resolution. The spectra were dumped to disk every second. Scans taken with calibration noise source turned on and off were used for amplitude calibration. The spectrum obtained after averaging the 4 transitions is shown in figure. A 3rd order polynomial was subtracted from the final spectrum to remove the residual baseline. 

The B0329+54 pulsar was observed with the 12m telescope at the center frequency of 3.1 GHz (see Fig.~\ref{fig:fig7}c). The 2.78-3.48 GHz filter was selected for these observations. Five Mock spectrometers were used to cover the 2.78-3.48 GHz bandwidth. Each spectrometer covered 172 MHz with 256 spectral channels. The center frequencies of the spectrometers were adjusted so that there will be overlaps in frequency between adjacent spectrometers. Spectral values from the overlapping frequency ranges were used to scale the power levels before combining the data from the different spectrometers. The data were processed using the \texttt{AO12PSR} pipeline. The resulting bandwidth after combining the data was 666 MHz and there were 992 spectral channels. The pulsar detection shown in Fig.~\ref{fig:fig7}c was obtained from a 30 minutes long observations.



The image of the Sun at X-band (8.647 GHz) made with the new receiver is shown in Fig.~\ref{fig:fig7}(d). The white circle on the image indicates the optical disk of the Sun. The observations were made on 07 August 2023 and the setup and observing procedure are same as those described in \citeA{mano23}. The near-simultaneous EUV images of the Sun observed respectively by the Atmospheric Imaging Assembly (AIA) in the wavelength band of 19.3 nm and the photospheric magnetogram recorded by the Helioseismic Magnetic Imager (HMI) on board the Solar Dynamics Observatory (SDO) \cite{pesnell2012}; \cite{lemen2012SoPh} are also shown in Fig.~\ref{fig:fig7}(d) for comparison. The bright emitting region in the north-west quadrant of the solar disk corresponds to an active region of complex magnetic configuration, which produced intense flares and coronal mass ejections.

Following the initial commissioning observations, the 12m telescope was used for shared risk observations, which included a survey of CH line emission from the Galactic center region and nearby molecular clouds, radio recombination line observations toward the Galactic center and Cygnus X region, HCCCN line observations toward galactic HII regions and continuation of pulsars monitoring program. Other observations that were being tried out with the new receiver include possible stellar flares and interplanetary scintillation studies. For most of these observations, undergraduate students from the ENCANTO (Enhancing and Nurturing Careers in Astronomy with New Training Opportunities) program\footnote{NSF Award Number:2219150; PI: Anish Roshi; Co-PIs: Brett Isham, Carmen Pantoja, Mayra Lebron, Abel Mendez} were actively participating and helping with observations, data processing and helping with identifying sources of RFI.

\section{Future Development Work}
\label{sec:future}

Further development work is required to improve the performance of the wideband system in four areas: (1) LNA dynamic range; (2) aperture efficiency; (3) full frequency coverage up to 14 GHz and (4) developing RFI robust warm receivers and backends. The need for these improvements and our suggestions are briefly described below.

The cryogenic receiver was designed to have an operating dynamic range of 20 dB at the first stage LNA. As discussed in Section~\ref{sec:wbrec}, the LNA output power is dominated by power due to RFI. Our initial RFI measurements showed that introducing a high-pass filter with cutoff frequency 2.5 GHz and -20 dB attenuation at 2.37 GHz could provide the desired dynamic range. However, the stop band attenuation of the high-pass filter degraded after cooling the filter to 15 K (see Fig.~\ref{fig:fig2} right). This filter degradation has limited the LNA operating dynamic range to 10 dB. Developing a cryogenic high-pass filter with the required characteristics can improve the LNA dynamic range. The increased dynamic range will enable, for example, Solar imaging at X-band \cite<see>[]{mano23} with the new receiver.

The aperture efficiency measured at 8.6 GHz is 44\%. Understanding the low aperture efficiency requires simulating the wideband feed response and developing a physical optics model for the 12m telescope. The primary and secondary reflectors of the 12m telescope are shaped reflectors and hence their geometry needs to be obtained for a complete simulation study. Such a study could provide insight on improving the aperture efficiency.

The current UDC operates up to a maximum input frequency of 12 GHz. An upgraded UDC was developed and tested at AO. This new UDC needs to be integrated with the receiver system to extend the operation of the 12m telescope up to 14 GHz. Currently the system is characterized only using the 2.78-3.48 GHz and 8.05-9.2 GHz filters in the filter bank. We will need to characterize the system with the other available filters.

In-band RFI is a major issue for the operation of the 12m telescope with the wideband receiver (see Fig.~\ref{fig:fig2} right). Increasing the dynamic range of LNA will certainly help with dealing with the in-band RFI (e.g. lower intermodulation products). In the future, we envisage designing and developing an RFI robust warm receiver system and backend with digitization done close to the front-end. This backend could also implement realtime RFI excision techniques. The 12m telescope could also be used for experimentation with active RFI mitigation techniques by implementing them in the digital hardware system.

\section{Summary}
\label{sec:sum}

In this paper, we presented the construction of a 2.5-14 GHz, cryogenic receiver. The receiver was successfully commissioned on the Arecibo 12m telescope in mid June 2023. The receiver was developed using a Quadruple-Ridged Flared Horn \cite<QRHF;>[]{Akgiray13}. In order to achieve sufficient dynamic range, we installed a highpass filter before the LNA to suppress strong RFI at frequencies below 2.7 GHz, which limited the lowest operating frequency to 2.5 GHz. The QRHF, highpass filter, noise couple and LNA are located inside a cryostat and are cooled to 15 K. The receiver temperature measured at the outdoor antenna facility was 25 K (median value) over 2.5 GHz to 14 GHz. The zenith system temperature measured on the 12m telescope was about 40 K near 3.1 and 8.6 GHz. The measured zenith telescope gains were 0.025 and 0.018 K/Jy respectively at 3.1 and 8.6 GHz. The estimated aperture efficiency at the two frequencies were 61\% and 44\% respectively. Future development includes  upgrading the highpass filter to improve RFI rejection near 2.5 GHz thus enhancing the operating dynamic range of the first stage LNA to 20 dB, improving the aperture efficiency at 8.6 GHz and upgrading the IF system to increase the upper frequency of operation from 12 GHz to 14 GHz.

\acknowledgments
The cryogenic system development for the 12m telescope was funded through the ``Restoration of Scientific Capabilities at the Arecibo Observatory after the impact of Hurricane Maria'' award (NSF Award Number:1930772). AO thanks Cosmic Microwave Technology Inc. for the donation of two cryogenic LNAs (Citcryo1-12D) for the project. We acknowledge extensive discussions with Sandy Weinreb, Caltech and Ganesh Rajagopalan, MIT, Haystack Observatory, during the initial design phase of the project, which helped in finalizing the cryogenic system configuration. We thank Carlos Rios, Wilson Morales, Francisco Jimenez, Osvaldo Colon and the maintenance staff for their efforts even under the difficult situation at the observatory. We also thank  Dr. Julie Brisset, PI of Arecibo Operations and Maintenance, and Dr. Olga Figueroa, Arecibo Observatory Director, for their support. A number of researchers from the Arecibo user community have contributed by suggesting science cases for the construction of the cryogenic system; we acknowledge their contribution.

\bibliography{12mcryo} 

\end{document}